\documentclass{article}
\usepackage{spconf,amsmath,graphicx}
\usepackage{amsfonts,amssymb}
\usepackage{float}
\usepackage{array}
\usepackage{booktabs}
\usepackage{epstopdf}

\title{VISinger: Variational Inference with adversarial learning for end-to-end SINGING VOICE SYNTHESIS}
\name{Yongmao Zhang$^1$, Jian Cong$^1$, Heyang Xue$^1$, Lei Xie$^1$$^*$, Pengcheng Zhu$^2$, Mengxiao Bi$^2$}
\address{
  $^1$Audio, Speech and Language Processing Group (ASLP@NPU)\\School of Computer Science,
  Northwestern Polytechnical University, Xi’an, China\\
  $^2$Fuxi AI Lab, NetEase Inc., Hangzhou, China}

\begin{document}
\ninept
\maketitle
\begin{abstract}

In this paper, we propose \textit{VISinger}, a complete end-to-end high-quality singing voice synthesis (SVS) system that directly generates singing audio from lyrics and musical score. Our approach is inspired by VITS~\cite{kim2021conditional}, an end-to-end speech generation model which adopts VAE-based posterior encoder augmented with normalizing flow based prior encoder and adversarial decoder. VISinger follows the main architecture of VITS, but makes substantial improvements to the prior encoder according to the characteristics of singing. First, instead of using phoneme-level mean and variance of acoustic features, we introduce a length regulator and a frame prior network to get the frame-level mean and variance on acoustic features, modeling the rich acoustic variation in singing. Second, we further introduce an F0 predictor to guide the frame prior network, leading to stabler singing performance. Finally, to improve the singing rhythm, we modify the duration predictor to specifically predict the phoneme to note duration ratio, helped with singing note normalization. Experiments on a professional Mandarin singing corpus show that VISinger significantly outperforms FastSpeech+Neural-Vocoder two-stage approach and the oracle VITS; ablation study demonstrates the effectiveness of different contributions.

\end{abstract}

\renewcommand{\thefootnote}{\fnsymbol{footnote}}
\footnotetext{* Corresponding author.}

\vspace{-8pt}
\section{Introduction}
\label{sec:intro}
\vspace{-5pt}
Automatic human voice generation has been significantly advanced by deep learning.
Aiming to mimic human speaking, text to speech (TTS) has witnessed tremendous progress with near human-parity performance. On the other hand, singing voice synthesis (SVS), which aims to generate singing voice from lyrics and music scores, has also been advanced with similar neural modeling framework in speech generation. Compared with speech synthesis, synthetic singing should not only be pronounced correctly according to the lyrics, but also conform to the labels of the music score. As the variation of acoustic features including fundamental frequency in singing is more abundant, and there are subtle pronunciation skills such as vibrato, modeling of singing voice is more challenging.

A typical deep learning based two-stage singing synthesis system generally consists of an acoustic model and a vocoder~\cite{ren2020deepsinger,Zhang2020DurIANSC,blaauw2020sequence,hono2018recent}. The acoustic model generates intermediate acoustic features from lyrics and music scores while the vocoder converts these acoustic features into waveform. For example, in~\cite{lu2020xiaoicesing, kim2018korean}, the neural acoustic model generates spectrum, excitation and aperiodicity parameters and then singing voice is synthesized using the World~\cite{morise2016world} vocoder. FastSpeech~\cite{ren2019fastspeech}, as a non-AR (auto-aggressive) model, was first used for speech generation and recently adopted in singing synthesis~\cite{lu2020xiaoicesing} with state-of-the-art performance. As neural vocoders, e.g., WaveRNN~\cite{kalchbrenner2018efficient} and HiFiGAN~\cite{kong2020hifi} have achieved high-fidelity speech generation, they have become the mainstream in current singing voice synthesis systems to reconstruct singing waveform from intermediate acoustic representation like mel-spectrum~\cite{chen2020hifisinger,lee2021n,gu2021bytesing}. 

Although these two-stage human voice generation systems made huge progress, the independent training of the two stages, i.e., the neural acoustic model and the neural vocoder, also leads to a \textit{mismatch} between the training and inference stages, resulting in degraded performance. Specifically, the neural vocoder is trained using the ground truth intermediate acoustic representation, e.g., mel-spectrum, but the predicted representation from the acoustic model is adopted during inference, resulting in distributional difference between the real and predicted intermediate representations. There are some tricks to alleviate this mismatch problem, including adopting the predicted acoustic features in neural vocoder fine-tuning and adversarial training~\cite{hono2019singing}.

A straightforward solution is to plug the two stages to become a unified model trained in an end-to-end manner. In text-to-speech synthesis, this kind of solutions have been recently explored, including FastSpeech2s~\cite{ren2020fastspeech}, EATS~\cite{donahue2020end}, Glow-WaveGAN~\cite{cong2021glow} and VITS~\cite{kim2021conditional}. In general, these works merge acoustic model and neural vocoder into one model enabling end-to-end learning or adopt a new latent representation instead of mel-spectrum to more easily confine the two parts work on the same distribution. Theoretically, end-to-end training can achieve better sound quality and simpler training and inference process. Among the end-to-end models, VITS uses a variational autoencoder (VAE)~\cite{kingma2013auto} to connect the acoustic model and the vocoder, which adopts variational inference augmented with normalizing flows and an adversarial training process, generating more natural-sounding than current two-stage models. 

In this paper, we build upon VITS and propose \textit{VISinger}, an end-to-end singing voice synthesis system based on variational inference (VI). To the best of our knowledge, VISinger is the first end-to-end solution in solving the two-stage mismatch problem in singing generation. It is non-trivial to adopt VITS in singing voice synthesis, because singing has substantial difference with speaking, although they both evolve from the same human vocal system. First, phoneme-level mean and variance of acoustic features are adopted in the flow-based prior encoder of VITS. In the singing task, we introduce a length regulator and a frame prior network to obtain the frame-level mean and variance instead, modeling the rich acoustic variation in singing and leading to more natural singing performance. As an ablation study, we find that simply increasing the number of layers of flow without adding the frame prior network can not achieve the same performance gain. Second, as intonational rendering is vital in singing, we particularly model the intonational aspects by introducing an F0 predictor to further guide the frame prior network, leading to more stable singing with natural intonation. Finally, to improve the rhythm delivery in singing, we modify the duration predictor to specifically predict the phoneme to note duration ratio, helped with singing note normalization. Experiments on a professional Mandarin singing corpus show that the proposed VISinger significantly outperforms the FastSpeech+Neural-Vocoder two-stage approach and the oracle VITS.

\vspace{-5pt}
\section{Method}
\label{sec:pagestyle}
\vspace{-5pt}

\begin{figure}[ht]
	\centering
	\includegraphics[scale=0.25]{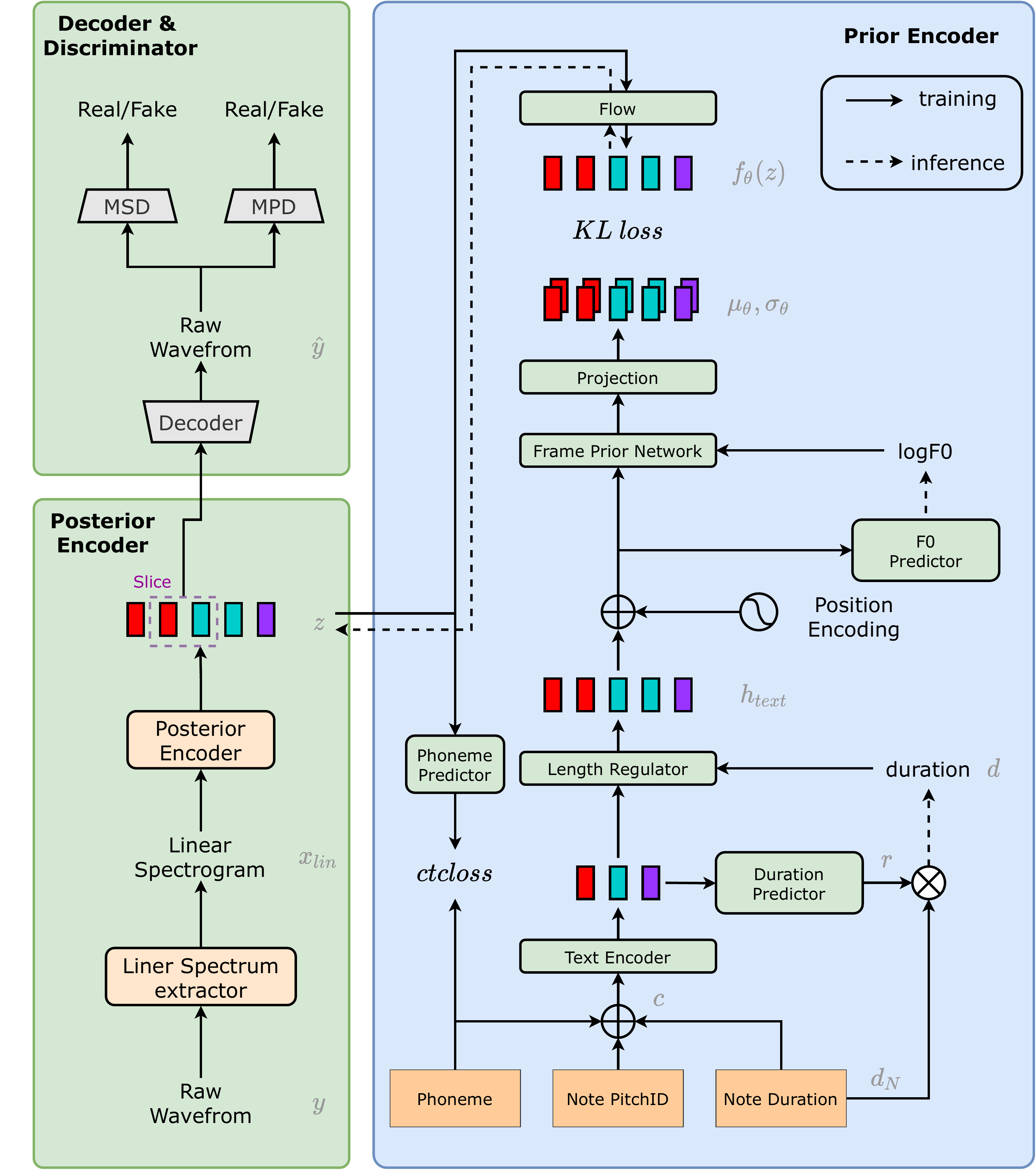}
	\caption{
		Architecture of VISinger
	}
	\label{model}
\end{figure}

As illustrated in Figure~\ref{model}, inspired by VITS~\cite{kim2021conditional}, we formulate the proposed model as a conditional variational autoencoder (CVAE), which mainly includes three parts: a posterior encoder, a prior encoder and a decoder. 
The posterior encoder extracts the latent representation $z$ from the waveform $y$, and the decoder reconstructs the waveform $\hat{y}$ according to $z$:
\begin{equation}
  \setlength{\abovedisplayskip}{3pt}
  \setlength{\belowdisplayskip}{3pt}
  z = Enc(y) \sim q(z|y)
  \label{eq_encoder}
\end{equation}
\begin{equation}
  \setlength{\abovedisplayskip}{3pt}
  \setlength{\belowdisplayskip}{3pt}
  \hat{y} = Dec(z) \sim p(y|z)
  \label{eq_2}
\end{equation}
In addition, we use a prior encoder to get a prior distribution $p(z|c)$ of the latent variables $z$ given music score condition $c$. CVAE adopts a reconstruction objective $L_{recon}$ and a prior regularization term as:
\begin{equation}
  \setlength{\abovedisplayskip}{3pt}
  \setlength{\belowdisplayskip}{3pt}
  L_{cvae} = L_{recon} + D_{KL}(q(z|y)||p(z|c)) + L_{ctc},
  \label{eq_vae_loss}
\end{equation}
where $D_{KL}$ is the Kullback-Leibler divergence and $L_{ctc}$ is the connectionist temporal classification (CTC) loss~\cite{CTC}. For the reconstruction loss, we use L1 distance of mel-spectrum between the ground truth and the generated waveform. In the following, we will introduce the details of the three modules.

\vspace{-15pt}
\subsection{Posterior Encoder}
\label{sec2:Posterior Encoder}
\vspace{-5pt}

The posterior encoder encodes the waveform $y$ into a latent representation $z$. To keep the original viewpoint in autoencoder, we treat the Liner Spectrum extractor as a fixed signal processing layer in the encoder. The encoder firstly transforms raw waveform to liner-spectrum with the signal processing layer. Similar with VITS, taking the linear spectrum as input, we use several WaveNet~\cite{oord2016wavenet} residual blocks to extract a sequence of hidden vector and then produces the mean and variance of the posterior distribution $p(z|y)$ by a linear projection. Then we can get the latent $z$ sampled from $p(z|y)$ using the reparametrization trick.

\subsection{Decoder}
\label{sec2:Decoder}
\vspace{-5pt}

The decoder generates audio waveform according to the extracted intermediate representation $z$. For more efficient training, we only feed the sliced $z$ into the decoder rather than the entire length to produce the corresponding audio segment. Following VITS, we also use GAN-based training to improve the quality of the reconstructed speech. The discriminator D follows HiFiGAN's Multi-Period Discriminator~(MPD) and Multi-Scale Discriminator~(MSD). Specifically, the GAN losses for the generator G and the discriminator D are defined as:
\begin{equation}
  \setlength{\abovedisplayskip}{3pt}
  \setlength{\belowdisplayskip}{3pt}
  L_{adv}(G)= \mathbb{E}_{(z)} \left[ (D(G(z))-1)^{2} \right] \label{LadvG}
\end{equation}
\begin{equation}
  \setlength{\abovedisplayskip}{3pt}
  \setlength{\belowdisplayskip}{3pt}
  L_{adv}(D)= \mathbb{E}_{(y,z)} \left[(D(y)-1)^{2}+(D(G(z)))^{2} \right]\label{LadvD}
\end{equation}
Furthermore, we use the feature matching loss $L_{fm}$ as an additional loss to learn the generator for more stable training. The feature matching loss minimizes the L1 distance between the feature map extracted from intermediate layers in each discriminator.

\subsection{Prior Encoder}
\label{sec2:Prior Encoder}
\vspace{-5pt}

Given music score condition $c$, the prior encoder provides the prior distribution ${p(z|c)}$ used for the prior regularization term of CVAE. The text encoder takes music score as input and produces a phoneme-level representation. To match the frequency of $z$, we use Length Regulator in FS2~\cite{ren2020fastspeech} to expand the phoneme-level representation to frame-level representation $h_{text}$. As acoustic variation is more salient in singing and different frames in a phoneme may obey different distributions, we add a frame prior network to generate fine-grained prior normal distribution with mean $\mu_{\theta}$ and variance $\sigma_{\theta}$.
In order to improve the expressiveness of prior distribution and provide more supervisory information for the potential variable~$z$, a normalizing flow $f_{\theta}$ and a phoneme predictor are added to the prior encoder. The phoneme predictor consists of two layers of FFT, and the output of the module will calculate the CTC loss~\cite{CTC} with phoneme.
\begin{equation}
  \setlength{\abovedisplayskip}{3pt}
  \setlength{\belowdisplayskip}{3pt}
  p(z|c) = N(f_\theta(z);\mu_\theta(c), \sigma_\theta(c))) \big \vert det \frac{\partial f_{\theta}(z)}{\partial_z } \big \vert .
\end{equation}

\vspace{-12pt}
\subsubsection{Text Encoder}
\label{sec2:Text Encoder}
\vspace{-5pt}

The music score of a song mainly includes lyrics, note duration and note pitch. We first convert the lyrics to a phoneme sequence. Note duration is the number of frames corresponding to each note, and note pitch is converted to Pitch ID following the MIDI standard~\cite{MIDI}. The note duration sequence and note pitch sequence are extended to the length of the phoneme sequence. 
The text encoder which consists of multiple FFT blocks takes the above three sequences in phoneme-level as input and generates a phoneme-level representation of music score.

\vspace{-5pt}
\subsubsection{Length Regulator}
\label{sec2:LR}
\vspace{-5pt}

Since the pronunciation of each phoneme in singing is rather complicated than speaking, we specifically add a Length Regulator~(LR) module to expand phoneme-level representation to frame-level representation $h_{text}$. In the training process, the ground truth duration $d$ corresponding to each phoneme is used for expansion, and the predicted duration $\hat{d}$ is used in the synthesis process. The duration predictor is composed of multiple one-dimensional convolution layers. Because the duration information in the score defines the overall rhythm and speed of singing, here we do not use the stochastic duration predictor as in VITS. Since the note duration $d_{N}$ in the music score provides a prior information for the duration prediction, we make the duration prediction based on the note duration. The duration predictor predicts the ratio of the phoneme duration to the corresponding note duration, denoted by $r$. We call this as note normalization~(Note Norm). Finally, the duration loss can be expressed as below:
\begin{equation}
  \setlength{\abovedisplayskip}{3pt}
  \setlength{\belowdisplayskip}{3pt}
  L_{dur}= \left\| r \times d_{N} - d \right\|_{2}\label{Ldur}
\end{equation}
\begin{equation}
  \setlength{\abovedisplayskip}{3pt}
  \setlength{\belowdisplayskip}{3pt}
  \hat{d} = r \times d_{N}
\end{equation}

\noindent The product of $r$ and $d_{N}$ is the predicted number of frames $\hat{d}$, which guides LR in the synthesis process. The predicted phoneme duration $\hat{d}$ will match the label in the score.

\vspace{-5pt}
\subsubsection{Frame Prior Network with F0 Predictor}
\label{sec2:Frame prior network}
\vspace{-5pt}

In the training process of the VITS model~\cite{kim2021conditional}, the text encoder extracts phoneme-level text information, serving as the prior knowledge for the extraction of the latent variable $z$. However, in the singing task, the variation of acoustic sequence in each phoneme is very rich, so it is not enough to represent the corresponding frame sequence only by using the mean and variance of phonemes. As a result, we find that there are many words with incorrect pronunciation in the audio synthesized by VITS. Therefore, one of the key improvements we propose is to add a frame prior network to the model. The frame prior network is composed of multiple layers of one-dimensional convolution. We found that simply increasing the number of layers of the flow model cannot have the same effect as adding this frame prior network module. Frame prior network performs post-processing on the frame-level sequence to obtain frame-level mean $\mu_{\theta}$ and variance $\sigma_{\theta}$. 

Since intonational rendering is vital in the naturalness of singing, we particularly introduce the F0 information to further guide the frame prior network, where F0 is obtained by an F0 predictor which consists of multiple FFT blocks during inference, and the LF0 loss is expressed as
\begin{small}
\begin{equation}
  \setlength{\abovedisplayskip}{3pt}
  \setlength{\belowdisplayskip}{3pt}
  L_{LF0}=\left\| \hat{LF0} - LF0 \right\|_{2}\label{LLF0}
\end{equation}
\end{small}
where $\hat{LF0}$ is the predicted LF0.

\vspace{-10pt}
\subsubsection{Flow}
\label{sec2:Flow}
\vspace{-5pt}

VISinger follows the flow decoder in VITS~\cite{kim2021conditional}. Flow is composed of multiple affine coupling layers~\cite{dinh2016density} to improve the flexibility of prior distribution. Flow can reversibly transform a normal distribution into a more general distribution, and the trained flow can realize the inverse transformation. During the training process, the latent variable $z$ is transformed to $f(z)$ through flow. In the inference process, we reverse the direction of flow and convert the output of the frame prior network into the latent variable $\hat{z}$. 

\subsection{Final Loss}
\label{sec2:Final Loss}
\vspace{-5pt}
With the above CVAE and adversarial training, we optimize our proposed model with the full objective:

\begin{small}
\begin{equation}
  \setlength{\abovedisplayskip}{3pt}
  \setlength{\belowdisplayskip}{3pt}
\begin{split}
    L=&L_{adv}(G)+L_{fm}(G)+ L_{cvae} + \lambda L_{dur}+\beta L_{LF0}
\end{split}
\end{equation}
\begin{equation}
  \setlength{\abovedisplayskip}{3pt}
  \setlength{\belowdisplayskip}{3pt}
    L(D)=L_{adv}(D)
\end{equation}
\end{small}
where $L_{adv}(G)$ and $L_{adv}(D)$ are the GAN loss of G and D respectively, and feature matching loss $L_{fm}$ is added to improve the stability of the training. The loss of CVAE $L_{cvae}$ consists of the reconstruction loss, the KL loss and the CTC loss.

\vspace{-5pt}
\section{Experiments}
\label{sec:exp}
\vspace{-5pt}

\subsection{Datasets}
\label{sec3:Datasets}
\vspace{-5pt}

Singing from a female singer is recorded in a professional studio, resulting in a 4.7 hours singing dataset with 100 songs. All the audios are recorded at 48kHz with 16-bit quantization, and we down-sampled them to 24k in our experiments. The phoneme duration is manually labeled by professional annotators. To facilitate model training, audio is split into segments of about 5 seconds, resulting in a total of 2616 segments. Among them, 2496 segments are randomly selected for training, while 16 and 104 segments are for validation and test.

\vspace{-5pt}
\subsection{Model Configuration}
\label{sec3:Model Configuration}
\vspace{-5pt}

We train the following models for comparison.
\begin{itemize}
  \item 
  \textbf{FastSpeech+Neural vocoder}: a typical two-stage singing synthesis system constructed by FastSpeech and neural vocoder where HiFiGAN and WaveRNN are compared.
  \item
  \textbf{VITS}: the singing synthesis model constructed by the original VITS.
  \item
  \textbf{VISinger}: the proposed system based on the VITS model, adopting all the contributions introduced in the paper.
\end{itemize}

Because of the complicated F0 contour in the singing task, in~\cite{lu2020xiaoicesing}, the residual link of F0 is added to make the decoder only need to predict the deviation on the note pitch. We refer to this idea to build the FastSpeech model and add the residual link from the note pitch embedding to the input of the decoder to adapt to the singing synthesis task. The dimensions of phoneme embedding, pitch embedding, and duration embedding are 256, 128, and 128, respectively. The rest of the hyperparameter settings are consistent with those in~\cite{ren2019fastspeech}. The generator version of the HiFiGAN vocoder we use is V1. The dimension of the GRU of WaveRNN is 512.

\begin{table*}[tb]
	\centering
	\caption{Experimental results in terms of subjective mean opinion score~(MOS) and two objective metrics.}
	\begin{tabular}{m{5cm}m{2.3cm}<{\centering}m{2.3cm}<{\centering}m{2.3cm}<{\centering}m{1.4cm}<{\centering}m{1.5cm}<{\centering}}
		\toprule
		Model&Model Size (M)&Sound quality&Naturalness&F0 MAE&Dur MAE  \\ 
		\midrule
		FastSpeech+HiFiGAN&52.47&3.80~($\pm$0.10)&3.73~($\pm$0.12)&13.90&9.94  \\
		FastSpeech+HiFiGAN \ (Fine-tuned) &52.47&3.85~($\pm$0.11)&3.75~($\pm$0.13)&13.74&9.94  \\
		FastSpeech+WaveRNN&41.58&3.78~($\pm$0.13)&3.74~($\pm$0.12)&13.96&9.94 \\
		FastSpeech+WaveRNN (Fine-tuned) &41.58&3.82~($\pm$0.16)&\textbf{3.79~($\pm$0.13)}&13.89&9.94 \\
		VITS&27.41&3.41~($\pm$0.15)&3.10~($\pm$0.17)&15.34&6.55  \\
		VISinger (Proposed)&36.50 &\textbf{3.93~($\pm$0.12)}&3.76~($\pm$0.10)&\textbf{13.69}&\textbf{4.73}  \\
		\midrule
		Recording &-& 4.38~($\pm$0.10)&4.32~($\pm$0.12)&-&-  \\
		\bottomrule
	\end{tabular}
	\vspace{-5pt}
	\label{MOS}
\end{table*}


In the proposed model, the embedding dimension of the input features is consistent with the baseline. The text encoder contains 6 FFT blocks, and the duration predictor consists of 3 one-dimensional convolutional networks. And note normalization is adopted in the duration predictor. The F0 predictor consists of 6 FFT blocks. The frame prior network is composed of 4 FFT blocks. The rest of the hyperparameter settings are consistent with those in VITS.

FastSpeech is trained up to 400k steps on a 2080Ti GPU with batch size of 32. HiFiGAN and WaveRNN are trained up to 800k steps with batch size of 32 as well. We train VITS and VISinger on 4 2080Ti GPUs. The batch size of each GPU is 8, and the two models are trained up to 600k steps. We also fine-tune the neural-vocoders to 100k steps with the predicted outputs from the first stage model.

\vspace{-5pt}
\subsection{Experimental Results}
\label{sec3:Evaluate}
\vspace{-5pt}

We performed a MOS test to evaluate the four systems. Specifically, we randomly selected 30 from 104 test segments for subjective listening and asked 10 listeners to assess the quality and naturalness. Objective metrics, including F0 mean absolute error~(F0 MAE) and duration mean absolute error~(Dur MAE), are also calculated. The F0 is extracted from the generated audio. The results are summarized in Table~\ref{MOS}.

\begin{figure}[htb]
	\centering
	\includegraphics[scale=0.11]{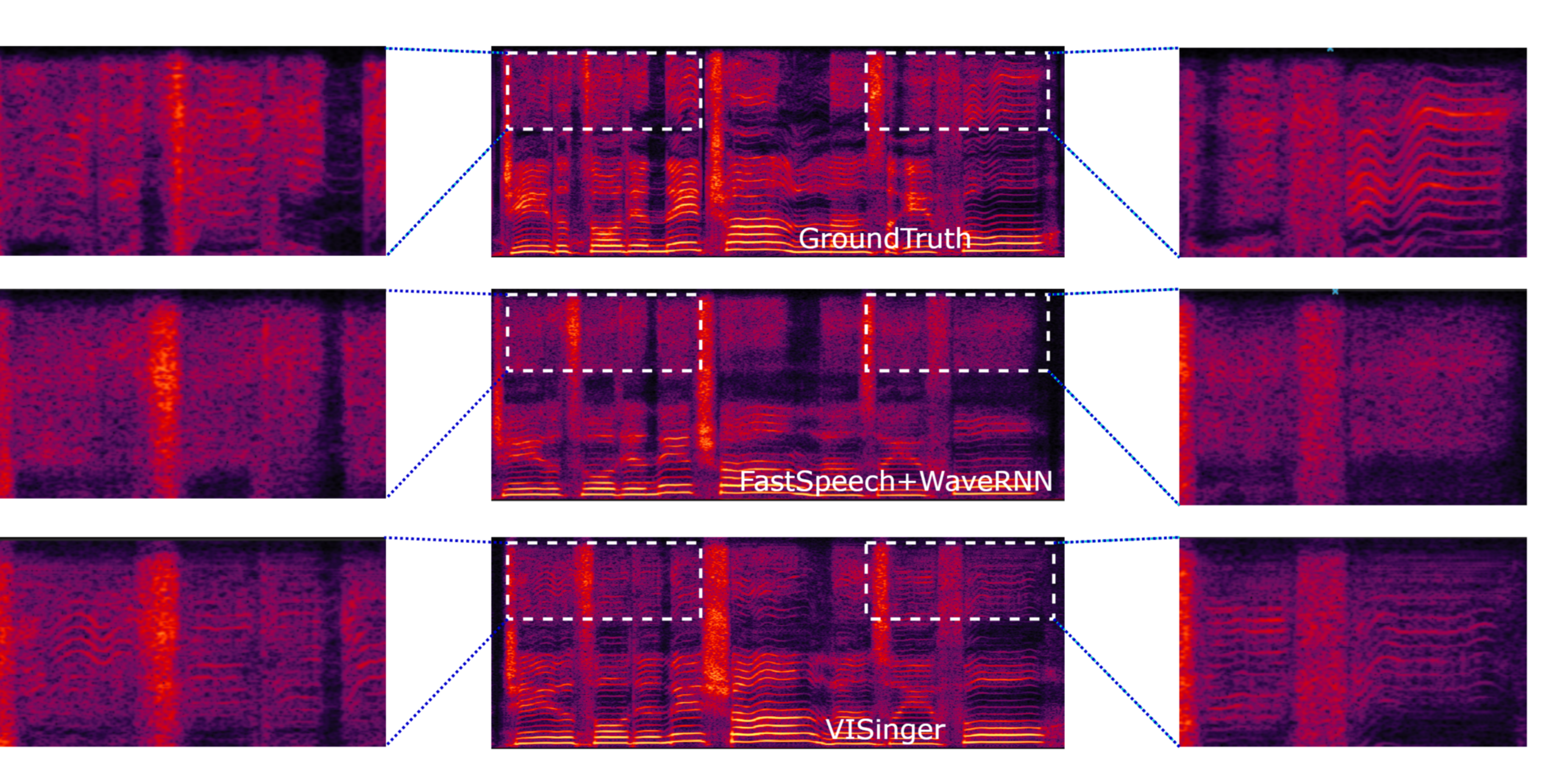}
	\caption{
		Spectrogram of a testing clip, generated by the ground truth, FastSpeech+WaveRNN, and VISinger.
	}
	\label{mels}
\end{figure}

Results in Table~\ref{MOS} show that VISinger has the best sound quality, F0 MAE and Dur MAE and its naturalness is comparative with that of FastSpeech+WaveRNN. Another advantage of an end-to-end model is model size: VISinger is smaller than the two-stage FastSpeech+WaveRNN model. The oracle VITS does not perform well and our substantial improvements on it lead such an end-to-end approach to the top in singing performance. We further visualize the spectrum of testing clips and Figure~\ref{mels} shows an example, where we can clearly see that the harmonic components in high frequency on the spectrum generated by VISinger is much clearer, which is close to the ground truth spectrum. By contrast, the harmonic components are not clear and more blurred in the high frequency area of the spectrum generated by FastSpeech+WaveRNN. According to the listeners, the singing generated by VISinger is more stable in intonation, without obvious pronunciation problems. However, the singing generated by VITS may have some intelligibility problems.



\subsection{Ablation study}
\label{sec3:Ablation study}
\vspace{-5pt}

We further conduct an ablation study to validate different contributions. We remove phoneme predictor, F0 predictor and frame prior network in turn and let the listeners assess the quality and naturalness. From Table~\ref{MOS for ablation study}, we can see that the MOS scores in quality and naturalness are both degraded with the removal of different components. To further show the contribution from the frame prior network, we simply double the layers of the flow model (from 4 to 8) in VITS. We find that this cannot lead to performance gain. Or we can conclude that the use of frame prior network is necessary. Note that the performance of VISinger is worse than VITS after deleting the F0 predictor and frame prior network. Because the ground truth duration used by LR may be biased during training, VISinger is less effective than VITS with Monotonic Alignment Search~(MAS)~\cite{kim2020glow} after deleting the frame prior network and F0 predictor.

\begin{table}[htb]
    
	\centering
	\caption{Ablation study with MOS test.}
	\begin{tabular}{lll}
		\toprule
		Model & Sound quality&Naturalness  \\ 
		\midrule
		VITS (flow 4 $\rightarrow$ 8) &3.52~($\pm$0.13) &3.19~($\pm$0.14)  \\
		VISinger (Proposed)                &\textbf{3.93~($\pm$0.12)}  &\textbf{3.76~($\pm$0.10)}  \\
        \ -~Phoneme Predictor  &3.92~($\pm$0.11)  &3.69~($\pm$0.12)  \\
        \ \ -~F0 predictor        &3.80~($\pm$0.10)  &3.58~($\pm$0.13)  \\
        \ \ \ -~Frame Prior Network &3.30~($\pm$0.14)  &2.78~($\pm$0.12) \\
		\bottomrule
	\end{tabular}
	\label{MOS for ablation study}
\end{table}

\begin{figure}[htb]
	\centering
	\includegraphics[scale=0.33]{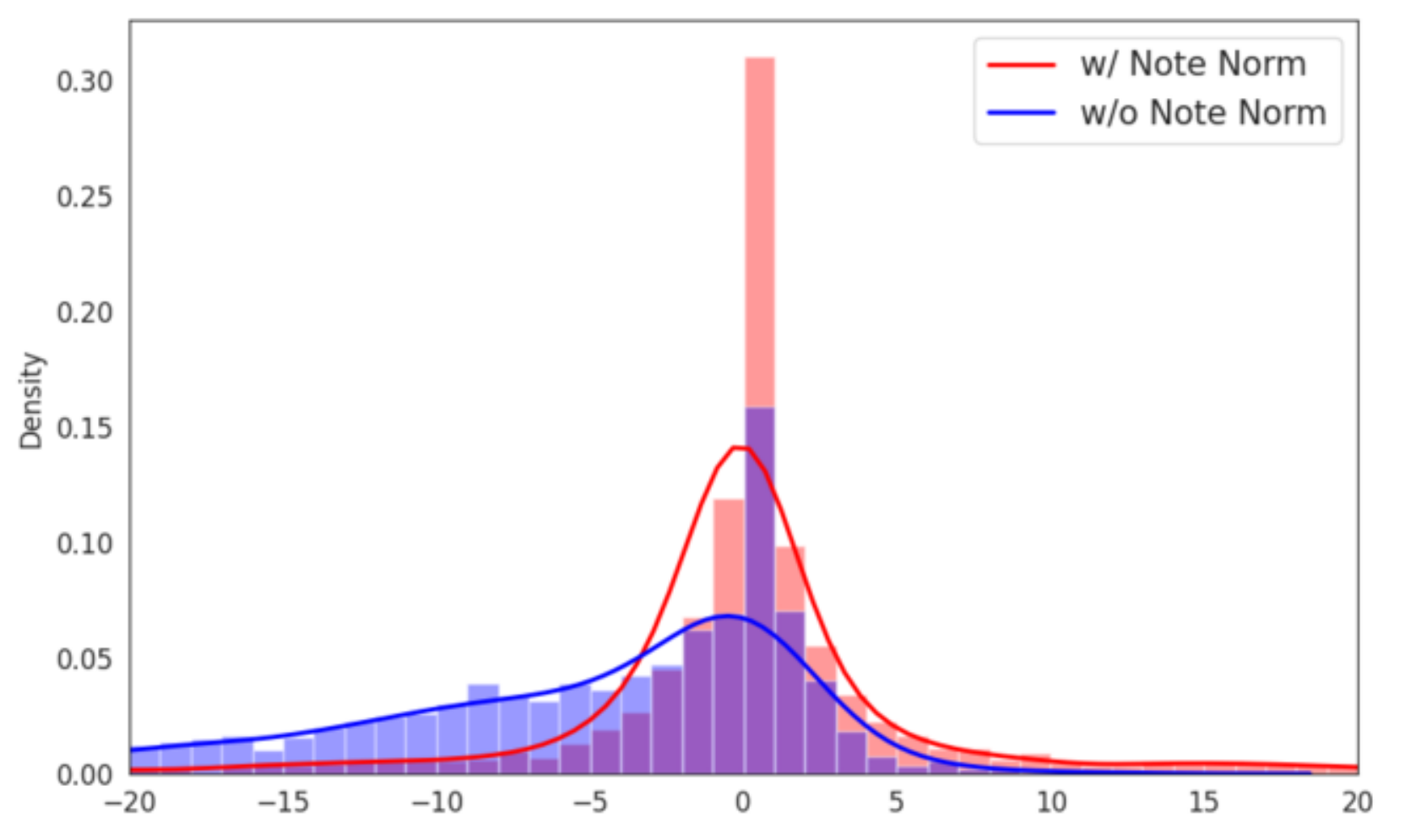}
	\caption{
		Duration deviation comparison for ablation study.
	}
	\label{duration deviation}
	\vspace{-5pt}
\end{figure}

To study the role of the note normalization strategy, we count the phoneme-level duration difference between the predicted and the ground truth in the test set, and the duration deviation is shown in Figure \ref{duration deviation}. In the figure, the horizontal axis represents the duration difference between the predicted and the ground truth, and the vertical axis indicates the frequency of the occurrence (density). Zero on the horizontal axis indicates that the prediction is absolutely accurate and smaller is better. We can see that, with the note normalization strategy, we have more correctly predicted duration values and the variance is not big. But without the note normalization strategy, we see more deviation in duration. Hence we can conclude that the note norm trick is quite effective.

\section{Conclusions}
\label{sec:Conclusions}
\vspace{-5pt}

In this paper, we proposed VISinger, a complete end-to-end singing synthesis system, to migrate the mismatch problem in conventional two-stage systems. The proposed system is built upon VITS -- the recently proposed end-to-end framework for speech generation. To let the framework work in singing voice generation, we proposed several contributions, including length regulator, frame prior network and modified duration predictor with note normalization. Experiments on a professional Mandarin singing corpus show that VISinger significantly outperforms the FastSpeech+Neural-Vocoder two-stage approach and the oracle VITS. In the future, we will continue to improve the end-to-end framework with the aim to close the performance gap between human singing and synthetic singing. We suggest the readers visit our demo page~$^1$.

\renewcommand{\thefootnote}{\fnsymbol{footnote}}
\footnotetext{$^1$~https://zhangyongmao.github.io/VISinger/}

\vfill\pagebreak

\bibliographystyle{IEEE}
\bibliography{strings,refs}

\begin{thebibliography}{10}

\bibitem{kim2021conditional}
Jaehyeon Kim, Jungil Kong, and Juhee Son,
\newblock ``Conditional variational autoencoder with adversarial learning for
  end-to-end text-to-speech,''
\newblock in {\em Proceedings of the 38th International Conference on Machine
  Learning}, 2021.

\bibitem{ren2020deepsinger}
Yi~Ren, Xu~Tan, Tao Qin, Jian Luan, Zhou Zhao, and Tie-Yan Liu,
\newblock ``Deepsinger: Singing voice synthesis with data mined from the web,''
\newblock in {\em Proceedings of the 26th ACM SIGKDD International Conference
  on Knowledge Discovery \& Data Mining}, 2020.

\bibitem{Zhang2020DurIANSC}
Liqiang Zhang, Chengzhu Yu, Heng Lu, Chao Weng, Chunlei Zhang, Yusong Wu, Xiang
  Xie, Zijin Li, and Dong Yu,
\newblock ``{DurIAN-SC: Duration Informed Attention Network Based Singing Voice
  Conversion System},''
\newblock in {\em Proc. Interspeech}, 2020.

\bibitem{blaauw2020sequence}
Merlijn Blaauw and Jordi Bonada,
\newblock ``Sequence-to-sequence singing synthesis using the feed-forward
  transformer,''
\newblock in {\em 2020 IEEE International Conference on Acoustics, Speech and
  Signal Processing (ICASSP)}, 2020.

\bibitem{hono2018recent}
Yukiya Hono, Shumma Murata, Kazuhiro Nakamura, Kei Hashimoto, Keiichiro Oura,
  Yoshihiko Nankaku, and Keiichi Tokuda,
\newblock ``Recent development of the dnn-based singing voice synthesis
  system—sinsy,''
\newblock in {\em Proceedings, APSIPA Annual Summit and Conference}, 2018.

\bibitem{lu2020xiaoicesing}
Peiling Lu, Jie Wu, Jian Luan, Xu~Tan, and Li~Zhou,
\newblock ``{XiaoiceSing: A High-Quality and Integrated Singing Voice Synthesis
  System},''
\newblock in {\em Proc. Interspeech}, 2020.

\bibitem{kim2018korean}
Juntae Kim, Heejin Choi, Jinuk Park, Minsoo Hahn, Sangjin Kim, and Jong-Jin
  Kim,
\newblock ``Korean singing voice synthesis system based on an lstm recurrent
  neural network,''
\newblock in {\em Proc. Interspeech}, 2018.

\bibitem{morise2016world}
Masanori Morise, Fumiya Yokomori, and Kenji Ozawa,
\newblock ``World: a vocoder-based high-quality speech synthesis system for
  real-time applications,''
\newblock {\em IEICE TRANSACTIONS on Information and Systems}, vol. 99, no. 7,
  2016.

\bibitem{ren2019fastspeech}
Yi~Ren, Yangjun Ruan, Xu~Tan, Tao Qin, Sheng Zhao, Zhou Zhao, and Tie-Yan Liu,
\newblock ``Fastspeech: Fast, robust and controllable text to speech,''
\newblock in {\em Advances in Neural Information Processing Systems}, 2019.

\bibitem{kalchbrenner2018efficient}
Nal Kalchbrenner, Erich Elsen, Karen Simonyan, Seb Noury, Norman Casagrande,
  Edward Lockhart, Florian Stimberg, Aaron Oord, Sander Dieleman, and Koray
  Kavukcuoglu,
\newblock ``Efficient neural audio synthesis,''
\newblock in {\em International Conference on Machine Learning}, 2018.

\bibitem{kong2020hifi}
Jungil Kong, Jaehyeon Kim, and Jaekyoung Bae,
\newblock ``Hifi-gan: Generative adversarial networks for efficient and high
  fidelity speech synthesis,''
\newblock in {\em Advances in Neural Information Processing Systems}, 2020,
  vol.~33.

\bibitem{chen2020hifisinger}
Jiawei Chen, Xu~Tan, Jian Luan, Tao Qin, and Tie-Yan Liu,
\newblock ``Hifisinger: Towards high-fidelity neural singing voice synthesis,''
\newblock {\em arXiv preprint arXiv:2009.01776}, 2020.

\bibitem{lee2021n}
Gyeong-Hoon Lee, Tae-Woo Kim, Hanbin Bae, Min-Ji Lee, Young-Ik Kim, and
  Hoon-Young Cho,
\newblock ``N-singer: A non-autoregressive korean singing voice synthesis
  system for pronunciation enhancement,''
\newblock in {\em Proc. Interspeech}, 2021.

\bibitem{gu2021bytesing}
Yu~Gu, Xiang Yin, Yonghui Rao, Yuan Wan, Benlai Tang, Yang Zhang, Jitong Chen,
  Yuxuan Wang, and Zejun Ma,
\newblock ``Bytesing: A chinese singing voice synthesis system using duration
  allocated encoder-decoder acoustic models and wavernn vocoders,''
\newblock in {\em 2021 12th International Symposium on Chinese Spoken Language
  Processing (ISCSLP)}, 2021.

\bibitem{hono2019singing}
Yukiya Hono, Kei Hashimoto, Keiichiro Oura, Yoshihiko Nankaku, and Keiichi
  Tokuda,
\newblock ``Singing voice synthesis based on generative adversarial networks,''
\newblock in {\em 2019 IEEE International Conference on Acoustics, Speech and
  Signal Processing (ICASSP)}, 2019.

\bibitem{ren2020fastspeech}
Yi~Ren, Chenxu Hu, Xu~Tan, Tao Qin, Sheng Zhao, Zhou Zhao, and Tie-Yan Liu,
\newblock ``Fastspeech 2: Fast and high-quality end-to-end text to speech,''
\newblock in {\em International Conference on Learning Representations}, 2020.

\bibitem{donahue2020end}
Jeff Donahue, Sander Dieleman, Miko{\l}aj Bi{\'n}kowski, Erich Elsen, and Karen
  Simonyan,
\newblock ``End-to-end adversarial text-to-speech,''
\newblock {\em In International Conference on Learning Representations}, 2021.

\bibitem{cong2021glow}
Jian Cong, Shan Yang, Lei Xie, and Dan Su,
\newblock ``Glow-wavegan: Learning speech representations from gan-based
  variational auto-encoder for high fidelity flow-based speech synthesis,''
\newblock in {\em Proc. Interspeech}, 2021.

\bibitem{kingma2013auto}
Diederik~P Kingma and Max Welling,
\newblock ``Auto-encoding variational bayes,''
\newblock in {\em International Conference on Learning Representations}, 2014.

\bibitem{CTC}
F.~Gomez A.~Graves, S.~Fernandez and J.~Schmidhuber,
\newblock ``Connectionist temporal classification: labelling unsegmented
  sequence data with recurrent neural networks,''
\newblock in {\em Proceedings of the 23rd international conference on Machine
  learning 2006}.

\bibitem{oord2016wavenet}
Aaron van~den Oord, Sander Dieleman, Heiga Zen, Karen Simonyan, Oriol Vinyals,
  Alex Graves, Nal Kalchbrenner, Andrew Senior, and Koray Kavukcuoglu,
\newblock ``Wavenet: A generative model for raw audio,''
\newblock {\em SSW, 125}, 2016.

\bibitem{MIDI}
M.~M. Association,
\newblock ``Midi manufacturers association,''
\newblock {\em https://www.midi.org.}

\bibitem{dinh2016density}
Laurent Dinh, Jascha Sohl-Dickstein, and Samy Bengio,
\newblock ``Density estimation using real nvp,''
\newblock in {\em International Conference on Learning Representations}, 2017.

\bibitem{kim2020glow}
Jaehyeon Kim, Sungwon Kim, Jungil Kong, and Sungroh Yoon,
\newblock ``Glow-tts: A generative flow for text-to-speech via monotonic
  alignment search,''
\newblock in {\em Advances in Neural Information Processing Systems}, 2020,
  vol.~33.

\end{thebibliography}

\end{document}